\documentclass[preprint,superscriptaddress,nofootinbib]{revtex4}
  
  \usepackage{graphicx}
  \begin{document}
  \title{Study of $J/{\psi}$ ${\to}$ $D_{s,d}V$ decays with perturbative QCD approach}
  \author{Yueling Yang}
  \affiliation{Institute of Particle and Nuclear Physics,
              Henan Normal University, Xinxiang 453007, China}
  \author{Junfeng Sun}
  \affiliation{Institute of Particle and Nuclear Physics,
              Henan Normal University, Xinxiang 453007, China}
  \author{Jie Gao}
  \affiliation{Institute of Particle and Nuclear Physics,
              Henan Normal University, Xinxiang 453007, China}
  \author{Qin Chang}
  \affiliation{Institute of Particle and Nuclear Physics,
              Henan Normal University, Xinxiang 453007, China}
  \author{Jinshu Huang}
  \affiliation{College of Physics and Electronic Engineering,
              Nanyang Normal University, Nanyang 473061, China}
  \author{Gongru Lu}
  \affiliation{Institute of Particle and Nuclear Physics,
              Henan Normal University, Xinxiang 453007, China}

  \begin{abstract}
  Inspired by the recent measurements on two-body nonleptonic
  $J/{\psi}$ weak decay at BESIII, the charm-changing $J/{\psi}$
  ${\to}$ $D_{s,d}V$ weak decays are studied with perturbative
  QCD approach, where $V$ denotes ${\rho}$ and $K^{\ast}$ vector
  mesons. It is found that branching ratio for $J/{\psi}$
  ${\to}$ $D_{s}{\rho}$ decay can reach up to ${\cal O}(10^{-9})$,
  which is within the potential measurement capability of the
  future high-luminosity experiments.
  \end{abstract}
  \keywords{$J/{\psi}$ meson; weak decay; branching ratio; perturbative QCD}
  \pacs{13.25.Gv 12.39.St 14.40.Pq 14.65.Dw}
  \maketitle

  \section{Introduction}
  \label{sec01}
  The $J/{\psi}$ particle is bound state of $c\bar{c}$ pair with
  given quantum numbers $I^{G}J^{PC}$ $=$ $0^{-}1^{--}$ \cite{pdg}.
  Since its discovery in 1974 \cite{bnl,slac}, the $J/{\psi}$ meson
  is always a hot and active topic for particle physicists.
  The $c\bar{c}$ pair of the $J/{\psi}$ meson annihilate mainly
  into gluons, which provides a valuable resource to explore the
  properties of the quark-gluon coupling and the invisible gluons,
  to search for various glueballs and possible exotic hadrons.
  There are two hierarchies in the $J/{\psi}$ meson and other
  heavy quarkonium, one is dynamical energy scales responsible
  for production and decay interactions of particles, and the
  other is relative velocity of $c$ quark\footnotemark[1].
  The $J/{\psi}$ meson plays a prominent role in investigation
  of QCD dynamical.
  \footnotetext[1]{According to the power counting rules of
  nonrelativistic quantum chromodynamics (NRQCD)
  \cite{prd46,prd51,rmp77}, there are several distinct energy
  scales in charmonium, for example, typical three-momentum
  $m_{c}v$ and kinetic energy $m_{c}v^{2}/2$, where $v$ ${\ll}$
  $1$ is the typical relative velocity of heavy quark. Those
  energy scales satisfy a hierarchy relation $m_{c}$ ${\gg}$
  $m_{c}v$ ${\gg}$ $m_{c}v^{2}$.}

  A conspicuous property of the $J/{\psi}$ meson is its narrow
  decay width, only about 30 ppm\footnotemark[2] of its mass.
  The $J/{\psi}$ meson lies below the kinematic $D\bar{D}$ threshold.
  Its hadronic decay into light hadrons violates the phenomenological
  Okubo-Zweig-Iizuka rules \cite{o,z,i}.
  Besides the decay dominated by the strong and electromagnetic
  interactions, the $J/{\psi}$ can also decay via the weak
  interaction within the standard model.
  In this paper, we will study the $J/{\psi}$ ${\to}$ $D_{s,d}{\rho}$,
  $D_{s,d}K^{\ast}$ weak decays with perturbative QCD (pQCD) approach
  \cite{pqcd1,pqcd2,pqcd3}.
  \footnotetext[2]{ppm means percent per million, i.e. $10^{-6}$.}

  Experimentally, thanks to the good performance of CLEO-c, BES,
  LHCb, B-factories, and so on, plenty of $J/{\psi}$ data samples
  have been accumulated. Recently, the $J/{\psi}$ ${\to}$ $D_{s}{\rho}$,
  $D_{u}K^{\ast}$ weak decays have been searched for at BESIII using
  part of the available $J/{\psi}$ samples \cite{prd89.071101}.
  It is eagerly expected to have about $10^{10}$ $J/{\psi}$ samples
  at BESIII per year with the designed luminosity \cite{cpc36},
  and over $10^{10}$ prompt $J/{\psi}$ samples at LHCb per $fb^{-1}$
  data \cite{epjc71}, which offers opportunities to discover phenomena
  that have been previously overlooked because of statistical limitations.
  So a careful scrutiny of $J/{\psi}$ weak decays at high-luminosity
  dedicated experiments may be possible in the future.
  In particular, the ``flavor tag'' of a single charged $D$ meson
  from $J/{\psi}$ decay will precisely identify potential signal
  from massive background. In addition, an abnormal large production
  rate of single $D$ meson from $J/{\psi}$ decay would be a hint
  of new physics.

  Theoretically, the $J/{\psi}$ ${\to}$ $D_{q}V$ decay is, in fact,
  induced by $c$ ${\to}$ $q$ $+$ $W^{+}$ transition at quark level,
  where $q$ $=$ $s$ and $d$, the virtual $W^{+}$ boson materializes
  into a pair of quarks which then hadronizes into a vector
  meson $V$ $=$ ${\rho}$ and $K^{\ast}$.
  As it is well known, there must be the participation of
  strong interaction in nonleptonic $J/{\psi}$ weak decay,
  and $c$ quark mass is between perturbative and
  nonperturbative domain.
  In recent years, some QCD-inspired methods,
  such as pQCD approach \cite{pqcd1,pqcd2,pqcd3},
  QCD factorization approach \cite{qcdf1,qcdf2,qcdf3},
  soft and collinear effective theory \cite{scet1,scet2,scet3,scet4},
  have been fully formulated to explain nonleptonic $B$ decays.
  The $J/{\psi}$ ${\to}$ $D_{q}V$ decays have been investigated
  based on collinear approximation \cite{plb252,ijmpa14,ahep2013,ijmpa30}.
  In this paper, the $J/{\psi}$ ${\to}$ $D_{s,d}V$ decays will be
  restudied based on $k_{T}$ factorization.
  It is expected to glean new insights into factorization mechanism,
  nonperturbative dynamics, final state interactions, and so on,
  from nonleptonic $J/{\psi}$ weak decay.

  This paper is organized as follows.
  The theoretical framework and amplitudes for $J/{\psi}$ ${\to}$
  $D_{s,d}V$ decays are given in section \ref{sec02}, followed by
  numerical results and discussion in section \ref{sec03}.
  Finally, we summarize in the last section.

  \section{theoretical framework}
  \label{sec02}
  \subsection{The effective Hamiltonian}
  \label{sec0201}
  Theoretically, one usually uses the effective Hamiltonian
  to describe hadron weak decay, where hard contributions can
  be decently factorized based on operator product expansion
  and the renormalization group (RG) method.
  The effective Hamiltonian responsible for $J/{\psi}$ ${\to}$
  $D_{s,d}V$ decay could be written as \cite{9512380},
   \begin{equation}
  {\cal H}_{\rm eff}\ =\
   \frac{G_{F}}{\sqrt{2}}\,
   \sum\limits_{q_{1},q_{2}}
   V_{cq_{1}}V_{uq_{2}}^{\ast}\,
   \Big\{ C_{1}({\mu})\,Q_{1}({\mu})
         +C_{2}({\mu})\,Q_{2}({\mu})
   \Big\} + {\rm h.c.}
   \label{hamilton},
   \end{equation}
  where $G_{F}$ ${\simeq}$ $1.166{\times}10^{-5}\,\text{GeV}^{-2}$ \cite{pdg}
  is Fermi constant; $q_{1,2}$ $=$ $d$ and $s$.

  The Cabibbo-Kobayashi-Maskawa (CKM) factors are written as
  \begin{equation}
  \begin{array}{lcl}
  V_{cs}V_{ud}^{\ast}\ =\
   1-{\lambda}^{2}-\frac{1}{2}A^{2}{\lambda}^{4}
  +{\cal O}({\lambda}^{6}),
  &~~&
  \text{for}~J/{\psi} {\to} D_{s}{\rho}~\text{decay}
  \\
  V_{cs}V_{us}^{\ast}\ =\
   {\lambda}-\frac{1}{2}{\lambda}^{3}-\frac{1}{8}{\lambda}^{5}(1+4A^{2})
  +{\cal O}({\lambda}^{6}),
  &~~&
  \text{for}~J/{\psi} {\to} D_{s}K^{\ast}~\text{decay}
  \\
  V_{cd}V_{ud}^{\ast}\ =\
  -V_{cs}V_{us}^{\ast}-A^{2}{\lambda}^{5}({\varrho}+i{\eta})
  +{\cal O}({\lambda}^{6}),
  &~~&
  \text{for}~J/{\psi} {\to} D_{d}{\rho}~\text{decay}
  \\
  V_{cd}V_{us}^{\ast}\ =\
  -{\lambda}^{2}+{\cal O}({\lambda}^{6}),
  &~~&
  \text{for}~J/{\psi} {\to} D_{d}K^{\ast}~\text{decay}
  \end{array}
  \label{eq:vckm}
  \end{equation}
  where $A$, ${\lambda}$, ${\varrho}$, ${\eta}$ are the Wolfenstein
  parameters; ${\lambda}$ $=$ ${\sin}{\theta}_{c}$ ${\simeq}$ $0.2$
  \cite{pdg} and ${\theta}_{c}$ is the Cabibbo angle.
  It is clearly seen that the $J/{\psi}$ ${\to}$ $D_{s}{\rho}$ decay
  is favored by the CKM factor $V_{cs}V_{ud}^{\ast}$.

  The Wilson coefficients $C_{1,2}(\mu)$ summarize the
  physical contributions above the scales of ${\mu}$.
  They are calculated at scale of the $W$ boson mass
  ${\mu}$ ${\sim}$ ${\cal O}(m_{W})$ with perturbation
  theory, and then evolved to scale of the $c$ quark
  mass ${\mu}$ ${\sim}$ ${\cal O}(m_{c})$ with RG evolution
  function,
  \begin{equation}
  \vec{C}({\mu}) = U_{4}({\mu},m_{b})U_{5}(m_{b},m_{W})\vec{C}(m_{W})
  \label{ci},
  \end{equation}
  where $U_{f}({\mu}_{j},{\mu}_{i})$ is RG evolution
  matrix \cite{9512380}. The Wilson coefficients are
  independent of a particular process in the same role
  of universal gauge couplings. They have properly
  been evaluated to the next-to-leading order.

  Generally, the penguin contributions induced by flavor
  changing neutral current transitions are proportional
  to small Wilson coefficients relative to tree
  contributions.
  Besides, for $c$ quark decay, the penguin contributions
  are also severely suppressed by the CKM factors
  $V_{cd}V_{ud}^{\ast}$ $+$ $V_{cs}V_{us}^{\ast}$ $=$
  $-V_{cb}V_{ub}^{\ast}$ ${\sim}$ ${\cal O}({\lambda}^{5})$.
  Hence, only the tree operators related to $W$ emission
  contributions are considered here.
  The expressions of tree operators are
    \begin{eqnarray}
    Q_{1} &=&
  [ \bar{q}_{1,{\alpha}}{\gamma}_{\mu}(1-{\gamma}_{5})c_{\alpha} ]
  [ \bar{u}_{\beta} {\gamma}^{\mu}(1-{\gamma}_{5})q_{2,{\beta}} ]
    \label{q1}, \\
    Q_{2} &=&
  [ \bar{q}_{1,{\alpha}}{\gamma}_{\mu}(1-{\gamma}_{5})c_{\beta} ]
  [ \bar{u}_{\beta} {\gamma}^{\mu}(1-{\gamma}_{5})q_{2,{\alpha}} ]
    \label{q2},
    \end{eqnarray}
  where ${\alpha}$ and ${\beta}$ are color indices.

  The physical contributions below scales of ${\mu}$ are
  included in hadronic matrix elements (HME).
  Because of the participation of the strong interaction,
  the entanglement perturbative and nonperturbative effects,
  the inadequate comprehension of hadronization mechanism and
  low energy QCD behavior, HME is the most complicated and
  intractable part. To get the amplitude, one has to face
  directly the HME calculation.

  \subsection{Hadronic matrix elements}
  \label{sec0202}
  Phenomenologically, the simplest approximation is that HME
  is parameterized into the production of transition form
  factors and decay constant based on naive factorization
  (NF) scheme \cite{nf}.
  The NF treatment on HME deprives any physical mechanism that
  could illustrate strong phases and rescattering among
  participating hadrons, and loses the ${\mu}$ dependence
  of HME which must exist to cancel that of Wilson coefficients.
  So the Lepage-Brodsky hard scattering approach \cite{prd22}
  is usually used, and HME is generally expressed as the
  convolution of hard scattering kernel with distribution
  amplitudes (DAs), where DAs reflect nonperturbative contributions
  but are universal. The hard part is, in principle, perturbatively
  calculable as a power of series of coupling ${\alpha}_{s}$.
  To suppress the soft contributions and avoid the problem of the
  endpoint singularity from collinear assumption \cite{qcdf1,qcdf2,qcdf3},
  the transverse momentum of quarks are retained explicitly and the
  Sudakov factors are introduced for each of meson wave functions
  in evaluation of potentially infrared contributions with pQCD
  approach \cite{pqcd1,pqcd2,pqcd3}.
  Finally, a decay amplitude could be written as a convolution
  integral of three parts \cite{pqcd1,pqcd2,pqcd3}: the hard effects
  enclosed by the Wilson coefficients $C_{i}$, the rescattering
  kernel amplitudes ${\cal H}$, and process-independent wave
  functions ${\Phi}$,
  \begin{equation}
  {\int} dk\,
  C_{i}(t)\,{\cal H}(t,k)\,{\Phi}(k)\,e^{-S}
  \label{hadronic},
  \end{equation}
  where $k$ is the momentum of valence quarks, $t$ is a typical
  scale and $e^{-S}$ is a Sudakov factor.

  \subsection{Kinematic variables}
  \label{sec0203}
  In the center-of-mass frame of $J/{\psi}$ meson, the light-cone
  kinematic variables are defined as follows.
  \begin{equation}
  p_{J/{\psi}}\, =\, p_{1}\, =\, \frac{m_{1}}{\sqrt{2}}(1,1,0)
  \label{kine-p1},
  \end{equation}
  \begin{equation}
  p_{D}\, =\, p_{2}\, =\, (p_{2}^{+},p_{2}^{-},0)
  \label{kine-p2},
  \end{equation}
  \begin{equation}
  p_{V}\, =\, p_{3}\, =\, (p_{3}^{-},p_{3}^{+},0)
  \label{kine-p3},
  \end{equation}
  \begin{equation}
  k_{i}\, =\, x_{i}\,p_{i}+(0,0,\vec{k}_{iT})
  \label{kine-ki},
  \end{equation}
  \begin{equation}
  {\epsilon}_{i}^{\parallel}\, =\,
   \frac{p_{i}}{m_{i}}-\frac{m_{i}}{p_{i}{\cdot}n_{+}}n_{+}
  \label{kine-longe},
  \end{equation}
  \begin{equation}
  n_{+}=(1,0,0)
  \label{kine-null-plus},
  \end{equation}
  \begin{equation}
  n_{-}=(0,1,0)
  \label{kine-null-minus},
  \end{equation}
  \begin{equation}
  p_{i}^{\pm}\, =\, (E_{i}\,{\pm}\,p)/\sqrt{2}
  \label{kine-pipm},
  \end{equation}
  \begin{equation}
  s\, =\, 2\,p_{2}{\cdot}p_{3}
  \label{kine-s},
  \end{equation}
  \begin{equation}
  t\, =\, 2\,p_{1}{\cdot}p_{2}\, =\ 2\,m_{1}\,E_{2}
  \label{kine-t},
  \end{equation}
  \begin{equation}
  u\, =\, 2\,p_{1}{\cdot}p_{3}\, =\ 2\,m_{1}\,E_{3}
  \label{kine-u},
  \end{equation}
  \begin{equation}
  p = \frac{\sqrt{ [m_{1}^{2}-(m_{2}+m_{3})^{2}]\,[m_{1}^{2}-(m_{2}-m_{3})^{2}] }}{2\,m_{1}}
  \label{kine-pcm},
  \end{equation}
  where the subscript $i$ $=$ $1$, $2$, $3$ on variables, including
  polarization vector ${\epsilon}_{i}$, four dimensional momentum $p_{i}$,
  energy $E_{i}$ and mass $m_{i}$, correspond to initial $J/{\psi}$ meson,
  recoiled $D$ meson, emitted vector meson $V$ $=$ ${\rho}$ and $K^{\ast}$,
  respectively;
  $x_{i}$ and $k_{i}$ ($\vec{k}_{iT}$) denote the longitudinal momentum
  fraction and (transverse) momentum of valence quarks,
  respectively; $n_{+}$ is the plus null vector;
  $s$, $t$ and $u$ are Lorentz transformation scalars;
  $p$ is the common momentum of final states.
  The kinematic variables are displayed in Fig.\ref{feynman}(a).

  \subsection{Wave functions}
  \label{sec0204}
  Taking the convention of Ref. \cite{prd65,jhep0703},  HME of the diquark
  operators squeezed between the vacuum and meson state
  is defined as below.
  \begin{equation}
 {\langle}0{\vert}c_{i}(z)\bar{c}_{j}(0){\vert}
 {\psi}(p_{1},{\epsilon}_{1}^{\parallel}){\rangle}\,
 =\, \frac{f_{\psi}}{4}{\int}d^{4}k_{1}\,e^{-ik_{1}{\cdot}z}
  \Big\{ \!\!\not{\epsilon}_{1}^{\parallel} \Big[
   m_{1}\,{\phi}_{\psi}^{v}(k_{1})
  -\!\!\not{p}_{1}\, {\phi}_{\psi}^{t}(k_{1})
  \Big] \Big\}_{ji}
  \label{wave-ccl},
  \end{equation}
  \begin{equation}
 {\langle}0{\vert}c_{i}(z)\bar{c}_{j}(0){\vert}
 {\psi}(p_{1},{\epsilon}_{1}^{\perp}){\rangle}\,
 =\, \frac{f_{\psi}}{4}{\int}d^{4}k_{1}\,e^{-ik_{1}{\cdot}z}
  \Big\{ \!\!\not{\epsilon}_{1}^{\perp} \Big[
   m_{1}\,{\phi}_{\psi}^{V}(k_{1})
  -\!\!\not{p}_{1}\, {\phi}_{\psi}^{T}(k_{1})
  \Big] \Big\}_{ji}
  \label{wave-cct},
  \end{equation}
  \begin{equation}
 {\langle}D_{q}(p_{2}){\vert}\bar{q}_{i}(z)c_{j}(0){\vert}0{\rangle} =
  \frac{if_{D_{q}}}{4}{\int}d^{4}k_{2}\,e^{ik_{2}{\cdot}z}\,
  \Big\{ {\gamma}_{5}\Big[ \!\!\not{p}_{2}\,{\Phi}_{D}^{a}(k_{2})
  +m_{2}\,{\Phi}_{D}^{p}(k_{2})\Big] \Big\}_{ji}
  \label{wave-ds},
  \end{equation}
  \begin{eqnarray} & &
 {\langle}V(p_{3},{\epsilon}_{3}^{\parallel})
 {\vert}u_{i}(z)\bar{q}_{j}(0){\vert}0{\rangle}
  \nonumber \\ &=&
  \frac{1}{4}{\int}_{0}^{1}dk_{3}\,e^{ik_{3}{\cdot}z}
  \Big\{ \!\!\not{\epsilon}_{3}^{\parallel}\,
   m_{3}\,{\Phi}_{V}^{v}(k_{3})
  +\!\!\not{\epsilon}_{3}^{\parallel}
   \!\!\not{p}_{3}\, {\Phi}_{V}^{t}(k_{3})
  -m_{3}\,{\Phi}_{V}^{s}(k_{3}) \Big\}_{ji}
  \label{wave-rhol},
  \end{eqnarray}
  \begin{eqnarray} & &
 {\langle}V(p_{3},{\epsilon}_{3}^{{\perp}})
 {\vert}u_{i}(z)\bar{q}_{j}(0){\vert}0{\rangle}\ =\
  \frac{1}{4}{\int}_{0}^{1}dk_{3}\,e^{ik_{3}{\cdot}z}
  \Big\{ \!\!\not{\epsilon}_{3}^{{\perp}}\,
   m_{3}\,{\Phi}_{V}^{V}(k_{3})
  \nonumber \\ & &
  +\!\!\not{\epsilon}_{3}^{{\perp}}
   \!\!\not{p}_{3}\, {\Phi}_{V}^{T}(k_{3})
  +\frac{i\,m_{3}}{p_{3}{\cdot}n_{+}}
  {\varepsilon}_{{\mu}{\nu}{\alpha}{\beta}}\,
  {\gamma}_{5}\,{\gamma}^{\mu}\,{\epsilon}_{3}^{{\perp}{\nu}}\,
  p_{3}^{\alpha}\,n_{+}^{\beta}\,
  {\Phi}_{V}^{A}(k_{3}) \Big\}_{ji}
  \label{wave-rhot},
  \end{eqnarray}
  where $f_{\psi}$ and $f_{D_{q}}$ are decay constants;
  wave functions ${\Phi}_{{\psi},V}^{v,T}$ and ${\Phi}_{D}^{a}$
  are twist-2; wave functions ${\Phi}_{{\psi},V}^{t,s,V,A}$ and
  ${\Phi}_{D}^{p}$ are twist-3. For wave functions of light vector
  meson, only ${\Phi}_{V}^{v}$ and ${\Phi}_{V}^{V,A}$ are involved
  in decay amplitudes (see Appendix \ref{blocks}). Their
  expressions are \cite{prd65,jhep0703}:
  \begin{equation}
 {\phi}_{V}^{v}(x) = 6\,x\,\bar{x}\, \Big\{ 1+
  \sum\limits_{i=1} a_{i}^{V}\,C_{i}^{3/2}(t) \Big\}
  \label{da-rhov},
  \end{equation}
  \begin{equation}
 {\phi}_{V}^{V}(x) \, =\, \frac{3}{4}\, (1+t^{2})
  \label{wave-rhoV},
  \end{equation}
  \begin{equation}
 {\phi}_{V}^{A}(x) \, =\, \frac{3}{2}\, (-t)
  \label{wave-rhoA}.
  \end{equation}
  where $\bar{x}$ $=$ $1$ $-$ $x$ and $t$ $=$ $\bar{x}$ $-$ $x$;
  $a_{i}^{V}$ is nonperturbative Gegenbauer moment and
  corresponds to Gegenbauer polynomial $C_{i}^{3/2}(t)$.
  \begin{equation}
  C_{1}^{3/2}(t) = 3\,t,
  \quad
  C_{2}^{3/2}(t) = \frac{3}{2}\,(5\,t^{2}-1),
  \quad
  {\cdots}
  \label{polynomials}
  \end{equation}

  With the relation of $m_{J/{\psi}}$ ${\simeq}$ $2m_{c}$ and
  $m_{D_{q}}$ ${\simeq}$ $m_{c}$ $+$ $m_{q}$, it is suspected
  that the motion of valence quarks in $J/{\psi}$ and $D_{q}$
  mesons is nearly nonrelativistic. So their spectrum can be
  described with time-independent Schr\"{o}dinger equation.
  Suppose the interaction between valence quarks is an isotropic
  harmonic oscillator potential, the ground state eigenfunction
  with quantum numbers $nL$ $=$ $1S$ is expressed as:
   \begin{equation}
  {\phi}_{1S}(\vec{k})\
  {\sim}\ e^{-\vec{k}^{2}/2{\omega}^{2}}
   \label{wave-k},
   \end{equation}
  where parameter ${\omega}$ determines the average transverse
  momentum, ${\langle}1S{\vert}k^{2}_{T}{\vert}1S{\rangle}$
  $=$ ${\omega}^{2}$. By using the transformation \cite{xiao},
   \begin{equation}
   \vec{k}^{2}\ {\to}\ \frac{1}{4} \Big(
   \frac{\vec{k}_{T}^{2}+m_{q_{1}}^{2}}{x_{1}}
  +\frac{\vec{k}_{T}^{2}+m_{q_{2}}^{2}}{x_{2}} \Big)
   \label{wave-kt},
   \end{equation}
  then integrating out transverse momentum $k_{T}$ and
  combining with their asymptotic forms,
  finally, DAs for $J/{\psi}$ and $D$ mesons are written as
   \begin{equation}
  {\phi}_{\psi}^{v}(x) = {\phi}_{\psi}^{T}(x) = A\, x\,\bar{x}\,
  {\exp}\Big\{ -\frac{m_{c}^{2}}{8\,{\omega}_{1}^{2}\,x\,\bar{x}} \Big\}
   \label{wave-clv},
   \end{equation}
   \begin{equation}
  {\phi}_{\psi}^{t}(x) = B\, t^{2}\,
  {\exp}\Big\{ -\frac{m_{c}^{2}}{8\,{\omega}_{1}^{2}\,x\,\bar{x}} \Big\}
   \label{wave-clt},
   \end{equation}
   \begin{equation}
  {\phi}_{\psi}^{V}(x) = C\, (1+t^{2})\,
  {\exp}\Big\{ -\frac{m_{c}^{2}}{8\,{\omega}_{1}^{2}\,x\,\bar{x}} \Big\}
   \label{wave-ctv},
   \end{equation}
   \begin{equation}
  {\phi}_{D}^{a}(x) = D\, x\,\bar{x}\, {\exp}\Big\{
  -\frac{\bar{x}\,m_{q}^{2}+x\,m_{c}^{2}}
        {8\,{\omega}_{2}^{2}\,x\,\bar{x}} \Big\}
   \label{wave-dqa},
   \end{equation}
   \begin{equation}
  {\phi}_{D}^{p}(x) = E\, {\exp}\Big\{
  -\frac{\bar{x}\,m_{q}^{2}+x\,m_{c}^{2}}
        {8\,{\omega}_{2}^{2}\,x\,\bar{x}} \Big\}
   \label{wave-dqp},
   \end{equation}
   where ${\omega}_{i}$ $=$ $m_{i}{\alpha}_{s}$ according
   to the NRQCD power counting rules \cite{prd46}, coefficients
   of $A$, $B$, $C$, $D$, $E$ are determined by
   the normalization conditions,
   \begin{equation}
  {\int}_{0}^{1}dx\,{\phi}_{\psi}^{v,t,V,T}(x) =1
   \label{normal-01},
   \end{equation}
   \begin{equation}
   {\int}_{0}^{1}dx\,{\phi}_{D}^{a,p}(x)=1
   \label{normal-02}.
   \end{equation}

  Here, it should be pointed out that there are many wave function
  models for $D$ meson, for example, Eq.(30) in Ref. \cite{prd78lv}.
  The preferred one in Ref. \cite{prd78lv} is:
   \begin{equation}
  {\phi}_{D}(x,b) = 6\,x\bar{x}\,\Big\{1+C_{D}(1-2x)\Big\}
  {\exp}\Big\{ -\frac{1}{2}w^{2}b^{2} \Big\}
   \label{wave-dqw},
   \end{equation}
  where $C_{D}$ $=$ $0.4$ and $w$ $=$ $0.2$ GeV for $D_{s}$ meson;
  $C_{D}$ $=$ $0.5$ and $w$ $=$ $0.1$ GeV for $D_{d}$ meson.
  In addition, the same form of Eq.(\ref{wave-dqw}), without
  a distinction between twist-2 and twist-3, is used in many
  practical calculation.

  \begin{figure}[h]
  \includegraphics[width=0.95\textwidth,bb=80 570 525 720]{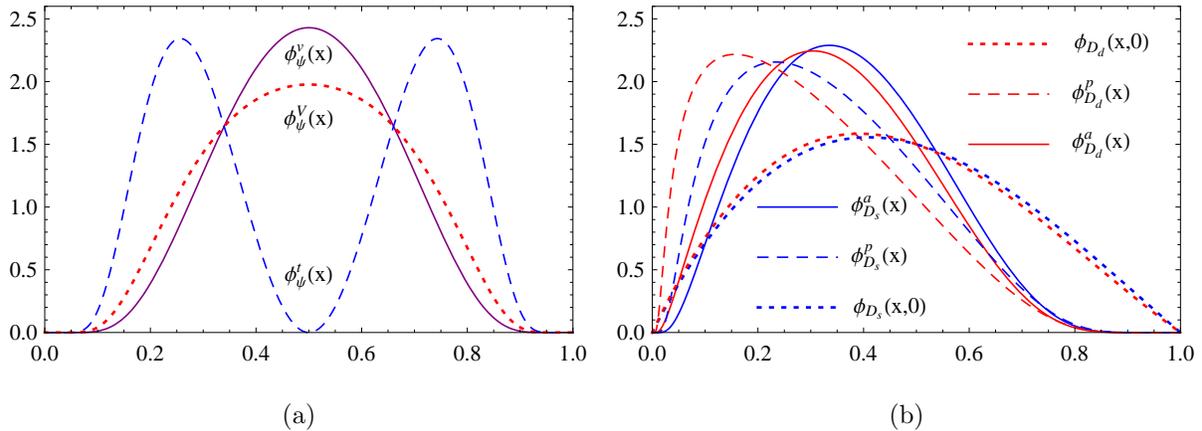}
  \caption{The shape lines of DAs for $J/{\psi}$ meson
  in (a) and $D_{d,s}$ mesons in (b), where
  ${\phi}_{\psi}^{v,t,V}(x)$, ${\phi}_{D}^{a,p}(x)$ and
  ${\phi}_{D}(x,b)$ correspond to Eqs.(\ref{wave-clv},\ref{wave-clt},\ref{wave-ctv}),
  Eqs.(\ref{wave-dqa},\ref{wave-dqp}) and Eq.(\ref{wave-dqw}), respectively.}
  \label{fig:wave}
  \end{figure}

  The shape lines of DAs for $J/{\psi}$ and $D_{s,d}$ mesons
  are displayed in Fig.\ref{fig:wave}. It is clearly seen that
  (1) DAs for $J/{\psi}$ meson is symmetric versus $x$, and a
  broad peak of ${\phi}_{D}^{a,p}(x)$ appears at $x$ $<$ $0.5$
  regions, which is basically in line with the picture that
  momentum fraction is proportional to valence quark mass.
  (2) under the influence of exponential functions, DAs of
  Eqs.(\ref{wave-clv}---\ref{wave-dqp}) fall quickly down to
  zero at endpoint $x$, $\bar{x}$ ${\to}$ $0$, which is bound
  to suppress soft contributions.
  (3) The flavor symmetry breaking effects between $D_{d}$ and
  $D_{s}$ mesons, and difference between twist-2 and twist-3
  are obvious in Eqs.(\ref{wave-dqa},\ref{wave-dqp})
  rather than Eq.(\ref{wave-dqw}). In this paper, we will use
  DAs of Eqs.(\ref{wave-dqa},\ref{wave-dqp}) for $D$ meson.

  \subsection{Decay amplitudes}
  \label{sec0205}
\begin{figure}[h]
  \includegraphics[width=0.99\textwidth,bb=75 620 530 725]{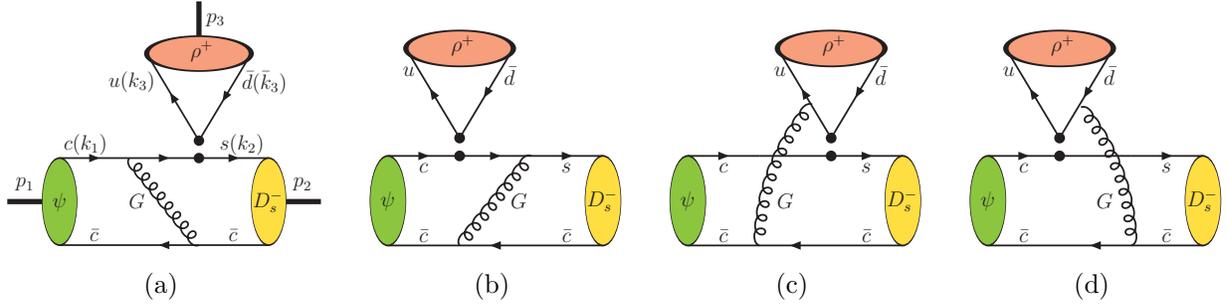}
  \caption{Feynman diagrams for $J/{\psi}$ ${\to}$ $D_{s}{\rho}$
   decay, including factorizable diagrams (a) and (b),
   and nonfactorizable diagrams (c) and (d).}
  \label{feynman}
  \end{figure}

  The Feynman diagrams for $J/{\psi}$ ${\to}$ $D_{s}{\rho}$
  decay are shown in Fig.\ref{feynman},
  including factorizable emission topologies (a) and (b)
  where gluon connects $J/{\psi}$ with $D_{s}$ meson,
  and nonfactorizable emission topologies (c) and (d)
  where gluon couples the spectator quark with
  emitted ${\rho}$ meson.

  The amplitude for $J/{\psi}$ ${\to}$ $D_{q}V$ decay is
  written as \cite{prd66}
   \begin{equation}
  {\cal A}(J/{\psi}{\to}D_{q}V) =
  {\cal A}_{L}({\epsilon}_{1}^{{\parallel}},{\epsilon}_{3}^{{\parallel}})
 +{\cal A}_{N}({\epsilon}_{1}^{{\perp}}{\cdot}{\epsilon}_{3}^{{\perp}})
 +i\,{\cal A}_{T}\,{\varepsilon}_{{\mu}{\nu}{\alpha}{\beta}}\,
  {\epsilon}_{1}^{{\mu}}\,{\epsilon}_{3}^{{\nu}}\,
   p_{1}^{\alpha}\,p_{3}^{\beta}
   \label{eq:amp01},
   \end{equation}
  which is conventionally written as helicity amplitudes \cite{prd66},
   \begin{equation}
  {\cal A}_{0}\ =\ -{\cal F}\,\sum\limits_{i}
  {\cal A}_{i,L}({\epsilon}_{1}^{{\parallel}},{\epsilon}_{3}^{{\parallel}})
   \label{eq:amp02},
   \end{equation}
   \begin{equation}
  {\cal A}_{\parallel}\ =\ \sqrt{2}\,{\cal F} \sum\limits_{i}
  {\cal A}_{i,N}
   \label{eq:amp03},
   \end{equation}
   \begin{equation}
  {\cal A}_{\perp}\ =\ \sqrt{2}\,{\cal F}\,m_{1}\,p \sum\limits_{i}
  {\cal A}_{i,T}
   \label{eq:amp04},
   \end{equation}
   \begin{equation}
  {\cal F}\ =\
  i\frac{G_{F}}{\sqrt{2}}\,
  \frac{C_{F}}{N_{c}}\,
  {\pi}\, f_{\psi}\,
  f_{D_{q}}\, f_{V}\,
  V_{cq_{1}} V_{uq_{2}}^{\ast}
   \label{eq:amp05},
   \end{equation}
  where the color number $N_{c}$ $=$ $3$ and color factor
  $C_{F}$ $=$ $(N_{c}^{2}-1)/2N_{c}$; the subscript $i$ on
  ${\cal A}_{i,j}$ corresponds to indices of Fig.\ref{feynman}.
  The expressions of building blocks ${\cal A}_{i,j}$
  can be found in Appendix \ref{blocks}.
  Our results show that (1) factorizable contributions
  [Fig.\ref{feynman} (a) and (b)] are color-favored, i.e.,
  $a_{1}$-dominated; (2) nonfactorizable contributions
  [Fig.\ref{feynman} (c) and (d)] are proportion to small
  Wilson coefficient $C_{2}$ and suppressed by color factor
  $1/N_{c}$.

  \section{Numerical results and discussion}
  \label{sec03}

  In the rest frame of $J/{\psi}$ meson, branching ratio is defined as
   \begin{equation}
  {\cal B}r\ = \frac{1}{12{\pi}}\,
   \frac{p}{m_{\psi}^{2}{\Gamma}_{\psi}}\, \Big\{
   {\vert}{\cal A}_{0}{\vert}^{2}
  +{\vert}{\cal A}_{\parallel}{\vert}^{2}
  +{\vert}{\cal A}_{\perp}{\vert}^{2} \Big\}
   \label{br}.
   \end{equation}

   \begin{table}[ht]
   \caption{The numerical values of input parameters.}
   \label{tab:input}
   \begin{ruledtabular}
   \begin{tabular}{lll}
   \multicolumn{3}{c}{CKM parameters\footnotemark[3] \cite{pdg}} \\ \hline
   \multicolumn{3}{c}{
    $A$          $=$ $0.814^{+0.023}_{-0.024}$, \qquad
    ${\lambda}$  $=$ $0.22537{\pm}0.00061$, \qquad
    $\bar{\varrho}$ $=$ $0.117{\pm}0.021$, \qquad
    $\bar{\eta}$ $=$ $0.353{\pm}0.013$,} \\ \hline
    \multicolumn{3}{c}{mass and decay constants} \\ \hline
    $m_{\psi}$  $=$ $3096.916{\pm}0.011$ MeV \cite{pdg},
  & $m_{D_{s}}$ $=$ $1968.30{\pm}0.11$ MeV \cite{pdg},
  & $m_{D_{d}}$ $=$ $1869.61{\pm}0.10$ MeV \cite{pdg}, \\
    $f_{\psi}$ $=$ $395.1{\pm}5.0$ MeV\footnotemark[4],
  & $f_{D_{s}}$ $=$ $257.5{\pm}4.6$ MeV \cite{pdg},
  & $f_{D_{d}}$ $=$ $204.6{\pm}5.0$ MeV \cite{pdg}, \\
    $m_{K^{\ast}}$ $=$ $891.66{\pm}0.26$ MeV \cite{pdg},
  & $m_{\rho}$ $=$ $775.26{\pm}0.25$ MeV \cite{pdg},
  & $m_{c}$ $=$ $1.67{\pm}0.07$ GeV \cite{pdg}, \\
    $f_{K^{\ast}}$ $=$ $220{\pm}5$ MeV \cite{jhep0703},
  & $f_{\rho}$ $=$ $216{\pm}3$ MeV \cite{jhep0703},
  & ${\Gamma}_{\psi}$ $=$ $92.9{\pm}2.8$ keV \cite{pdg}, \\
    $m_{s}$ ${\approx}$ $510$ MeV \cite{uds},
  & $m_{d}$ ${\approx}$ $310$ MeV \cite{uds}, \\ \hline
  \multicolumn{3}{c}{Gegenbauer moments at ${\mu}$ $=$ 1 GeV \cite{jhep0703}} \\ \hline
  \multicolumn{3}{c}{
    $a_{1}^{K^{\ast}}$ $=$ $-0.03{\pm}0.02$, \qquad
    $a_{2}^{K^{\ast}}$ $=$ $0.11{\pm}0.09$, \qquad
    $a_{1}^{\rho}$ $=$ $0$, \qquad
    $a_{2}^{\rho}$ $=$ $0.15{\pm}0.07$.}
  \end{tabular}
  \end{ruledtabular}
  \end{table}
  \footnotetext[3]{The relations between CKM parameters (${\varrho}$, ${\eta}$)
   and ($\bar{\varrho}$, $\bar{\eta}$) are \cite{pdg}: $({\varrho}+i{\eta})$ $=$
   $\displaystyle \frac{ \sqrt{1-A^{2}{\lambda}^{4}}(\bar{\varrho}+i\bar{\eta}) }
  { \sqrt{1-{\lambda}^{2}}[1-A^{2}{\lambda}^{4}(\bar{\varrho}+i\bar{\eta})] }$.}
  \footnotetext[4]{The decay constant $f_{\psi}$ can be obtained from
  experimental branching ratios for electromagnetic $J/{\psi}$ decay
  into charged lepton pairs through the formula
   \begin{equation}
  {\cal B}r(J/{\psi}{\to}{\ell}^{+}{\ell}^{-}) =
   \frac{16{\pi}}{27}f_{\psi}^{2}
   \frac{{\alpha}_{\rm QED}^{2}}{m_{\psi}\,{\Gamma}_{\psi}}
   \sqrt{ 1-4\frac{m_{\ell}^{2}}{m_{\psi}^{2}} }
   \Big\{ 1+2\frac{m_{\ell}^{2}}{m_{\psi}^{2}} \Big\}
   \label{eq:fjpsi},
   \end{equation}
  where ${\alpha}_{\rm QED}$ is the fine-structure constant,
  ${\ell}$ $=$ $e$ and ${\mu}$.
  One can get \cite{pdg} $f_{\psi}$ $=$ $395.4{\pm}7.0$ MeV with
  ${\cal B}r(J/{\psi}{\to}e^{+}e^{-})$ $=$ $(5.971{\pm}0.032)\%$,
  and $f_{\psi}$ $=$ $394.8{\pm}7.1$ MeV with
  ${\cal B}r(J/{\psi}{\to}{\mu}^{+}{\mu}^{-})$ $=$ $(5.961{\pm}0.033)\%$,
  respectively, where the errors arise from mass $m_{\psi}$, decay
  width ${\Gamma}_{\psi}$ and branching ratios.
  The weighted average is $f_{\psi}$ $=$ $395.1{\pm} 5.0 $ MeV.}

   \begin{table}[ht]
   \caption{Branching ratios for $J/{\psi}$ ${\to}$ $DV$ decays,
   where uncertainties of our results come from scale
   $(1{\pm}0.1)t_{i}$, quark mass $m_{c}$, hadronic parameters
   and CKM parameters, respectively.}
   \label{tab:output}
   \begin{ruledtabular}
   \begin{tabular}{lccccc}
   \multicolumn{1}{c}{Reference} & \cite{ijmpa14}\footnotemark[5] & \cite{ahep2013}
   & \cite{ijmpa30} & \cite{epjc55} & this work \\ \hline
    $10^{9}{\times}{\cal B}r(J/{\psi}{\to}D_{s}{\rho})$
  & $2.54$ & $5.1$ & $2.2$ & $1.3$
  & $3.33^{+0.97+0.47+0.17+0.002}_{-0.42-0.51-0.17-0.002}$ \\ \hline
    $10^{10}{\times}{\cal B}r(J/{\psi}{\to}D_{s}K^{\ast})$
  & $1.48$ & $2.8$ & $1.2$ & $0.8$
  & $1.86^{+0.57+0.28+0.12+0.010}_{-0.24-0.35-0.12-0.010}$  \\ \hline
    $10^{10}{\times}{\cal B}r(J/{\psi}{\to}D_{d}{\rho})$
  & $1.54$ & $2.2$ & $1.1$ & $0.4$
  & $1.32^{+0.37+0.14+0.08+0.007}_{-0.16-0.19-0.08-0.007}$ \\ \hline
    $10^{11}{\times}{\cal B}r(J/{\psi}{\to}D_{d}K^{\ast})$
  & ... & $1.3$ & $0.6$ & ...
  & $0.80^{+0.23+0.05+0.06+0.009}_{-0.10-0.12-0.06-0.009}$
  \end{tabular}
  \end{ruledtabular}
  \end{table}
  \footnotetext[5]{The updated results are listed in Table 4 of Ref. \cite{ahep2013}.}

  The values of input parameters are listed in Table \ref{tab:input},
  where if it is not specified explicitly, their central values will
  be taken as the default inputs.
  Our numerical results are presented in Table \ref{tab:output},
  where the first uncertainty comes from the choice of the typical
  scale $(1{\pm}0.1)t_{i}$, and expression of $t_{i}$ is
  given in Eq.(\ref{tab}) and Eq.(\ref{tcd});
  the second uncertainty is from quark mass $m_{c}$;
  the third uncertainty is from hadronic parameters including
  decay constants and Gegenbauer moments; and the fourth
  uncertainty of branching ratio comes from CKM parameters.
  The following are some comments.

  (1)
  As it is aforementioned, the $J/{\psi}$ decay modes considered
  here are dominated by the color-favored factorizable contributions
  and insensitive to nonfactorizable contributions. So, generally,
  branching ratio for a given $J/{\psi}$ ${\to}$ $D_{s,d}V$ decay
  has the same order of magnitude even with different
  phenomenological models.

  (2)
  There is a clear hierarchical pattern among branching ratios,
  mainly resulting from the hierarchical structure of CKM
  factors in Eq.(\ref{eq:vckm}), i.e.,
  \begin{equation}
  {\cal B}r(J/{\psi}{\to}D_{s}{\rho})\, {\gg}\,
  {\cal B}r(J/{\psi}{\to}D_{s}K^{\ast})\, {\sim}\,
  {\cal B}r(J/{\psi}{\to}D_{d}{\rho})\, {\gg}\,
  {\cal B}r(J/{\psi}{\to}D_{d}K^{\ast})
  \label{eq:br-4}.
  \end{equation}
  In addition, because nonfactorizable contributions are
  suppressed by both small $C_{2}$ and color factor $1/N_{c}$,
  there is an approximate relationship,
   \begin{equation}
   \frac{ {\cal B}r(J/{\psi}{\to}D_{s}K^{\ast}) }
        { {\cal B}r(J/{\psi}{\to}D_{s}{\rho}) }
   \ {\approx}\
   \frac{ {\cal B}r(J/{\psi}{\to}D_{d}K^{\ast}) }
        { {\cal B}r(J/{\psi}{\to}D_{d}{\rho}) }
   \ {\approx}\
  {\lambda}^{2}\,
   \frac{ f^{2}_{K^{\ast}} }
        { f^{2}_{\rho} }
   \label{eq:rbr}.
   \end{equation}
  Above all,  the Cabibbo- and color-favored $J/{\psi}$ ${\to}$
  $D_{s}{\rho}$ decay has branching ratio ${\sim}$ ${\cal O}(10^{-9})$,
  which is well within the measurement capability of the future
  high-luminosity experiments, such as super tau-charm factory,
  LHC and SuperKEKB.

  \begin{figure}[h]
  \includegraphics[width=0.55\textwidth,bb=105 475 505 720]{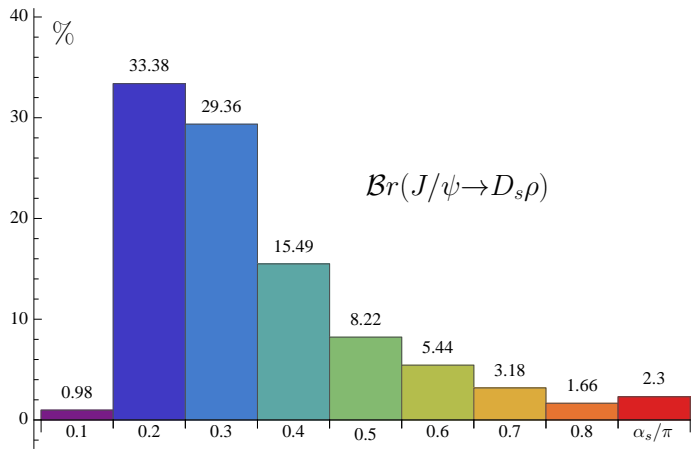}
  \caption{Contributions to branching ratio
  ${\cal B}r(J/{\psi}{\to}D_{s}{\rho})$ versus ${\alpha}_{s}/{\pi}$,
  where the numbers over histogram denote the percentage of the
  corresponding contributions.}
  \label{fig:as}
  \end{figure}

  (3)
  Here, one might question the practicability of pQCD approach and
  the feasibility of perturbative calculation because $c$ quark mass
  seems to be not large enough. To clear this issue up or to check what
  percentage of contributions come from perturbative domain,
  contributions to branching ratio ${\cal B}r(J/{\psi}{\to}D_{s}{\rho})$
  from different ${\alpha}_{s}/{\pi}$ region are displayed in Fig.\ref{fig:as}.
  It is easily seen that about 80\% contributions come from
  ${\alpha}_{s}/{\pi}$ ${\le}$ 0.4 regions, which implies that
  the calculation with pQCD approach is valid.
  One of crucial reasons for the small percentage in the region
  ${\alpha}_{s}/{\pi}$ ${\le}$ $0.1$ is that the absolute values
  of Wilson coefficients $C_{1,2}$, $a_{1}$ and coupling
  ${\alpha}_{s}$ decrease along with the increase of
  renormalization scale ${\mu}$. Of course, a perturbative
  calculation with pQCD approach is influenced by many factors,
  such as Sudakov factors, the choice of scale $t$, models of
  wave functions, etc., which deserve much attention but
  beyond the scope of this paper.

  (4)
  There are many uncertainties on branching ratios.
  The first uncertainty from scale $t$ could be reduced
  by the inclusion of higher order corrections to HME and an
  improved control on nonperturbative contributions.
  The second uncertainty from wave function models or parameter
  $m_{c}$ will be greatly lessened with the relative rate of
  branching ratios, for example, Eq.(\ref{eq:rbr}).
  The third uncertainty is dominated by decay constants
  whose effects will be weakened with the increasing
  precision of experimental measurements and/or theoretical
  calculation using nonperturbative methods (such as lattice
  QCD and so on). The uncertainty from CKM factor is small.
  Moreover, other factors, such as the final state
  interactions which is important and necessary for $c$
  quark decay, are not properly considered here, but deserve
  massive dedicated study.
  Our results just provide an order of magnitude estimation
  on branching ratio.

  \section{Summary}
  \label{sec04}
  Within the standard model, the $J/{\psi}$ meson can decay via
  the weak interaction, besides the strong and electromagnetic
  interactions. With anticipation of copious $J/{\psi}$ data samples
  at the future high-luminosity experiments and gradual improvement
  of particle identification techniques, we investigated the charm-changing
  $J/{\psi}$ ${\to}$ $D_{s,d}{\rho}$, $D_{s,d}K^{\ast}$ weak decays
  with pQCD approach.
  It is found that the estimated branching ratio for the color-
  and CKM-favored $J/{\psi}$ ${\to}$ $D_{s}{\rho}$ decay can be
  up to ${\cal O}(10^{-9})$, which is very likely to be measured
  in the future.

  \section*{Acknowledgments}
  We thank Professor Dongsheng Du (IHEP@CAS) and Professor Yadong
  Yang (CCNU) for helpful discussion.
  The work is supported by the National Natural Science Foundation
  of China (Grant Nos. 11547014, 11475055, 11275057 and U1332103).

  \begin{appendix}
  \section{Building blocks of decay amplitudes}
  \label{blocks}
  The expressions of building blocks ${\cal A}_{i,j}$
  are listed as follows, where subscript $i$ corresponds to
  indices of Fig.\ref{feynman}; and $j$ corresponds to
  helicity amplitudes.
   \begin{eqnarray}
  {\cal A}_{a,L} &=&
  {\int}_{0}^{1}dx_{1}
  {\int}_{0}^{1}dx_{2}
  {\int}_{0}^{\infty}b_{1}db_{1}
  {\int}_{0}^{\infty}b_{2}db_{2}\,
  {\phi}_{\psi}^{v}(x_{1})\,
  E_{a}(t_{a})
   \nonumber \\ &{\times}&
  H_{a}({\alpha},{\beta}_{a},b_{1},b_{2})\,
  {\alpha}_{s}(t_{a})\, a_{1}(t_{a})\,
   \Big\{ {\phi}_{D}^{p}(x_{2})\, m_{2}\,m_{c}\,u
   \nonumber \\ & &
   + {\phi}_{D}^{a}(x_{2})\, \Big[ m_{1}^{2}\,s-
   (4\,m_{1}^{2}\,p^{2}+m_{2}^{2}\,u)\,\bar{x}_{2}
   \Big] \Big\}
   \label{amp:al},
   \end{eqnarray}
   \begin{eqnarray}
  {\cal A}_{a,N} &=&
  m_{1}\, m_{3}
  {\int}_{0}^{1}dx_{1}
  {\int}_{0}^{1}dx_{2}
  {\int}_{0}^{\infty}b_{1}db_{1}
  {\int}_{0}^{\infty}b_{2}db_{2}\,
  {\phi}_{\psi}^{V}(x_{1})
   \nonumber \\ &{\times}&
  E_{a}(t_{a})\,
  H_{a}({\alpha},{\beta}_{a},b_{1},b_{2})\,
   \Big\{ -2\,m_{2}\,m_{c}\,{\phi}_{D}^{p}(x_{2})
   \nonumber \\ & &
   + {\phi}_{D}^{a}(x_{2})\, \Big[
   2\,m_{2}^{2}\,\bar{x}_{2} -t \Big]
   \Big\}\, {\alpha}_{s}(t_{a})\, a_{1}(t_{a})
   \label{amp:an},
   \end{eqnarray}
   \begin{eqnarray}
  {\cal A}_{a,T} &=&
  2\, m_{1}\,m_{3}
  {\int}_{0}^{1}dx_{1}
  {\int}_{0}^{1}dx_{2}
  {\int}_{0}^{\infty}b_{1}db_{1}
  {\int}_{0}^{\infty}b_{2}db_{2}\,
  {\phi}_{\psi}^{V}(x_{1})
  \nonumber \\ &{\times}&
  {\phi}_{D}^{a}(x_{2})\,
   E_{a}(t_{a})\,
   H_{a}({\alpha},{\beta}_{a},b_{1},b_{2})\,
   {\alpha}_{s}(t_{a})\,a_{1}(t_{a})
   \label{amp:at},
   \end{eqnarray}
   \begin{eqnarray}
  {\cal A}_{b,L} &=&
  {\int}_{0}^{1}dx_{1}
  {\int}_{0}^{1}dx_{2}
  {\int}_{0}^{\infty}b_{1}db_{1}
  {\int}_{0}^{\infty}b_{2}db_{2}\,
  H_{b}({\alpha},{\beta}_{b},b_{2},b_{1})
   \nonumber \\ &{\times}&
   E_{b}(t_{b})\,
   \Big\{ {\phi}_{\psi}^{v}(x_{1})\,{\phi}_{D}^{a}(x_{2})\,
   \Big[ m_{1}^{2}\, (s-4\,p^{2})\,\bar{x}_{1}-m_{2}^{2}\,u \Big]
   \nonumber \\ & &
  +2\,m_{1}\,m_{2}\,{\phi}_{\psi}^{t}(x_{1})\,
   {\phi}_{D}^{p}(x_{2})\, (s-u\,\bar{x}_{1}) \Big\}\,
   {\alpha}_{s}(t_{b})\, a_{1}(t_{b})
   \label{amp:bl},
   \end{eqnarray}
   \begin{eqnarray}
  {\cal A}_{b,N} &=&
  {\int}_{0}^{1}dx_{1}
  {\int}_{0}^{1}dx_{2}
  {\int}_{0}^{\infty}b_{1}db_{1}
  {\int}_{0}^{\infty}b_{2}db_{2}\,
   H_{b}({\alpha},{\beta}_{b},b_{2},b_{1})
   \nonumber \\ &{\times}&
   E_{b}(t_{b})\, {\alpha}_{s}(t_{b})\,
   \Big\{ m_{1}\,m_{3}\,{\phi}_{\psi}^{V}(x_{1})\,
   {\phi}_{D}^{a}(x_{2})\,(2\,m_{2}^{2}-t\,\bar{x}_{1})
   \nonumber \\ & &
   + 2\,m_{2}\,m_{3}\,{\phi}_{\psi}^{T}(x_{1})\,
   {\phi}_{D}^{p}(x_{2})\, (2\,m_{1}^{2}\,\bar{x}_{1}-t)
   \Big\}\, a_{1}(t_{b})
   \label{amp:bn},
   \end{eqnarray}
   \begin{eqnarray}
  {\cal A}_{b,T} &=& 2\, m_{3}
  {\int}_{0}^{1}dx_{1}
  {\int}_{0}^{1}dx_{2}
  {\int}_{0}^{\infty}b_{1}db_{1}
  {\int}_{0}^{\infty}b_{2}db_{2}\,
  H_{b}({\alpha}_{e},{\beta}_{b},b_{2},b_{1})\,
  E_{b}(t_{b})
   \nonumber \\ &{\times}&
  {\alpha}_{s}(t_{b})\,a_{1}(t_{b})\,
   \Big\{ 2\,m_{2}\,{\phi}_{\psi}^{T}(x_{1})\,{\phi}_{D}^{p}(x_{2})
  -m_{1}\,{\phi}_{\psi}^{V}(x_{1})\,{\phi}_{D}^{a}(x_{2})\, \bar{x}_{1} \Big\}
   \label{amp:bt},
   \end{eqnarray}
   \begin{eqnarray}
  {\cal A}_{c,L} &=&
   \frac{1}{N_{c}}
  {\int}_{0}^{1}dx_{1}
  {\int}_{0}^{1}dx_{2}
  {\int}_{0}^{1}dx_{3}
  {\int}_{0}^{\infty}db_{1}
  {\int}_{0}^{\infty}b_{2}db_{2}
  {\int}_{0}^{\infty}b_{3}db_{3}
   \nonumber \\ &{\times}&
  {\phi}_{\rho}^{v}(x_{3})\, E_{c}(t_{c})\,
   H_{c}({\alpha},{\beta}_{c},b_{2},b_{3})\,
   {\alpha}_{s}(t_{c})\, {\delta}(b_{1}-b_{2})
   \nonumber \\ &{\times}&
   C_{2}(t_{c})\,
   \Big\{ {\phi}_{\psi}^{v}(x_{1})\,{\phi}_{D}^{a}(x_{2})\,u\,
   \Big[ t\,(\bar{x}_{2}-\bar{x}_{1})+s\,(x_{3}-\bar{x}_{2})\Big]
   \nonumber \\ & &
   + {\phi}_{\psi}^{t}(x_{1})\, {\phi}_{D}^{p}(x_{2})\,m_{1}\,m_{2}\,
   \Big[ u\,(\bar{x}_{1}-x_{3})+s\,(x_{3}-\bar{x}_{2})\Big] \Big\}
   \label{amp:cl},
   \end{eqnarray}
   \begin{eqnarray}
  {\cal A}_{c,N} &=&
   \frac{ m_{3} }{N_{c}}
  {\int}_{0}^{1}dx_{1}
  {\int}_{0}^{1}dx_{2}
  {\int}_{0}^{1}dx_{3}
  {\int}_{0}^{\infty}db_{1}
  {\int}_{0}^{\infty}b_{2}db_{2}
  {\int}_{0}^{\infty}b_{3}db_{3}
   \nonumber \\ &{\times}&
   E_{c}(t_{c})\,
   H_{c}({\alpha},{\beta}_{c},b_{2},b_{3})\,
  {\alpha}_{s}(t_{c})\, C_{2}(t_{c})\,
  {\delta}(b_{1}-b_{2})
   \nonumber \\ &{\times}&
   \Big\{ {\phi}_{\psi}^{V}(x_{1})\,
  {\phi}_{D}^{a}(x_{2})\,
  {\phi}_{\rho}^{V}(x_{3})\,2\,m_{1}\,
   \Big[ s\,(\bar{x}_{2}-x_{3})+t\,(\bar{x}_{1}-\bar{x}_{2})\Big]
   \nonumber \\ & &
  +{\phi}_{\psi}^{T}(x_{1})\,
  {\phi}_{D}^{p}(x_{2})\,
  {\phi}_{\rho}^{V}(x_{3})\,m_{2}\,
   \Big[ u\,(x_{3}-\bar{x}_{1})+t\,(\bar{x}_{2}-\bar{x}_{1})\Big]
   \nonumber \\ & &
  +{\phi}_{\psi}^{T}(x_{1})\,
  {\phi}_{D}^{p}(x_{2})\,
  {\phi}_{\rho}^{A}(x_{3})\,2\,m_{1}\,m_{2}\,p\,(x_{3}-\bar{x}_{2}) \Big\}
   \label{amp:cn},
   \end{eqnarray}
   \begin{eqnarray}
  {\cal A}_{c,T} &=&
   \frac{ 1 }{N_{c}}
   \frac{ m_{3} }{m_{1}\,p}
  {\int}_{0}^{1}dx_{1}
  {\int}_{0}^{1}dx_{2}
  {\int}_{0}^{1}dx_{3}
  {\int}_{0}^{\infty}db_{1}
  {\int}_{0}^{\infty}b_{2}db_{2}
  {\int}_{0}^{\infty}b_{3}db_{3}
   \nonumber \\ &{\times}&
   E_{c}(t_{c})\,
   H_{c}({\alpha},{\beta}_{c},b_{2},b_{3})\,
  {\alpha}_{s}(t_{c})\, C_{2}(t_{c})\,
  {\delta}(b_{1}-b_{2})
   \nonumber \\ &{\times}&
   \Big\{ {\phi}_{\psi}^{V}(x_{1})\,
  {\phi}_{D}^{a}(x_{2})\,
  {\phi}_{\rho}^{A}(x_{3})\,2\,m_{1}\,
   \Big[ s\,(\bar{x}_{2}-x_{3})+t\,(\bar{x}_{1}-\bar{x}_{2})\Big]
   \nonumber \\ & &
  +{\phi}_{\psi}^{T}(x_{1})\,
  {\phi}_{D}^{p}(x_{2})\,
  {\phi}_{\rho}^{A}(x_{3})\,m_{2}\,
   \Big[ u\,(x_{3}-\bar{x}_{1})+t\,(\bar{x}_{2}-\bar{x}_{1})\Big]
   \nonumber \\ & &
  +{\phi}_{\psi}^{T}(x_{1})\,
  {\phi}_{D}^{p}(x_{2})\,
  {\phi}_{\rho}^{V}(x_{3})\,2\,m_{1}\,m_{2}\,p\,(x_{3}-\bar{x}_{2}) \Big\}
   \label{amp:ct},
   \end{eqnarray}
   \begin{eqnarray}
  {\cal A}_{d,L} &=&
   \frac{1}{N_{c}}
  {\int}_{0}^{1}dx_{1}
  {\int}_{0}^{1}dx_{2}
  {\int}_{0}^{1}dx_{3}
  {\int}_{0}^{\infty}db_{1}
  {\int}_{0}^{\infty}b_{2}db_{2}
  {\int}_{0}^{\infty}b_{3}db_{3}
   \nonumber \\ &{\times}&
  {\phi}_{\rho}^{v}(x_{3})\, E_{d}(t_{d})\,
   H_{d}({\alpha},{\beta}_{d},b_{2},b_{3})\,
  {\alpha}_{s}(t_{d})\,
  {\delta}(b_{1}-b_{2})
   \nonumber \\ &{\times}&
   \Big\{ {\phi}_{\psi}^{t}(x_{1})\,{\phi}_{D}^{p}(x_{2})\,m_{1}\,m_{2}\,
   \Big[ u\,(x_{3}-x_{1})+s\,(x_{2}-x_{3})\Big]
   \nonumber \\ & &
   + {\phi}_{\psi}^{v}(x_{1})\, {\phi}_{D}^{a}(x_{2})\,4\,m_{1}^{2}\,p^{2}\,
   (x_{3}-x_{2}) \Big\}\,  C_{2}(t_{d})
   \label{amp:dl},
   \end{eqnarray}
   \begin{eqnarray}
  {\cal A}_{d,N} &=&
  \frac{ m_{2}\,m_{3} }{N_{c}}
  {\int}_{0}^{1}dx_{1}
  {\int}_{0}^{1}dx_{2}
  {\int}_{0}^{1}dx_{3}
  {\int}_{0}^{\infty}db_{1}
  {\int}_{0}^{\infty}b_{2}db_{2}
  {\int}_{0}^{\infty}b_{3}db_{3}
   \nonumber \\ &{\times}&
  {\phi}_{\psi}^{T}(x_{1})\,
  {\phi}_{D}^{p}(x_{2})\,
  E_{d}(t_{d})\,
  H_{d}({\alpha},{\beta}_{d},b_{2},b_{3})\,
  {\alpha}_{s}(t_{d})\,  C_{2}(t_{d})\,
  {\delta}(b_{1}-b_{2})
   \nonumber \\ &{\times}&
   \Big\{ {\phi}_{\rho}^{V}(x_{3})\,
   \Big[ 2\,m_{1}^{2}\,x_{1}-t\,x_{2} -u\,x_{3}\Big]
   + 2\,m_{1}\,p\,{\phi}_{\rho}^{A}(x_{3})\,(x_{2}-x_{3}) \Big\}
   \label{amp:dn},
   \end{eqnarray}
   \begin{eqnarray}
  {\cal A}_{d,T} &=&
  \frac{ 1 }{N_{c}}\,
  \frac{ m_{2}\,m_{3} }{m_{1}\,p }
  {\int}_{0}^{1}dx_{1}
  {\int}_{0}^{1}dx_{2}
  {\int}_{0}^{1}dx_{3}
  {\int}_{0}^{\infty}db_{1}
  {\int}_{0}^{\infty}b_{2}db_{2}
  {\int}_{0}^{\infty}b_{3}db_{3}
   \nonumber \\ &{\times}&
  {\phi}_{\psi}^{T}(x_{1})\,
  {\phi}_{D}^{p}(x_{2})\,
  E_{d}(t_{d})\,
  H_{d}({\alpha},{\beta}_{d},b_{2},b_{3})\,
  {\alpha}_{s}(t_{d})\,  C_{2}(t_{d})\,
  {\delta}(b_{1}-b_{2})
   \nonumber \\ &{\times}&
   \Big\{ {\phi}_{\rho}^{A}(x_{3})\,
   \Big[ 2\,m_{1}^{2}\,x_{1}-t\,x_{2} -u\,x_{3}\Big]
   + 2\,m_{1}\,p\,{\phi}_{\rho}^{V}(x_{3})\,(x_{2}-x_{3}) \Big\}
   \label{amp:dt},
   \end{eqnarray}
  where $b_{i}$ is the conjugate variable of the transverse
  momentum $k_{iT}$; ${\alpha}_{s}$ is the QCD running coupling;
  $a_{1}$ $=$ $C_{1}$ $+$ $C_{2}/N_{c}$;
  $C_{1,2}$ are the Wilson coefficients.

  The hard scattering function $H_{i}$ and Sudakov factor $E_{i}$
  are defined as follows.
   \begin{equation}
   H_{a(b)}({\alpha},{\beta},b_{i},b_{j}) =
   K_{0}(b_{i}\sqrt{-{\alpha}})
   \Big\{ {\theta}(b_{i}-b_{j})
   K_{0}(b_{i}\sqrt{-{\beta}})
   I_{0}(b_{j}\sqrt{-{\beta}})
   + (b_{i}{\leftrightarrow}b_{j}) \Big\}
   \label{hab},
   \end{equation}
   \begin{eqnarray}
   H_{c(d)}({\alpha},{\beta},b_{2},b_{3}) &=&
   \Big\{ {\theta}(-{\beta}) K_{0}(b_{3}\sqrt{-{\beta}})
  +\frac{{\pi}}{2} {\theta}({\beta}) \Big[
   iJ_{0}(b_{3}\sqrt{{\beta}})
   -Y_{0}(b_{3}\sqrt{{\beta}}) \Big] \Big\}
   \nonumber \\ &{\times}&
   \Big\{ {\theta}(b_{2}-b_{3})
   K_{0}(b_{2}\sqrt{-{\alpha}})
   I_{0}(b_{3}\sqrt{-{\alpha}})
   + (b_{2}{\leftrightarrow}b_{3}) \Big\}
   \label{hcd},
   \end{eqnarray}
   \begin{equation}
   E_{i}(t)\ =\ \left\{ \begin{array}{ll}
   {\exp}\{ -S_{\psi}(t)-S_{D}(t) \}, & ~~\text{ for } i=a,b \\
   {\exp}\{ -S_{\psi}(t)-S_{D}(t)-S_{V}(t) \}, & ~~\text{ for } i=c,d
    \end{array} \right.
   \label{sudakov}
   \end{equation}
   \begin{equation}
   S_{\psi}(t)\ =\
   s(x_{1},p_{1}^{+},1/b_{1})
  +2{\int}_{1/b_{1}}^{t}\frac{d{\mu}}{\mu}{\gamma}_{q}
   \label{sudakov-cc},
   \end{equation}
   \begin{equation}
   S_{D}(t)\ =\
   s(x_{2},p_{2}^{+},1/b_{2})
  +2{\int}_{1/b_{2}}^{t}\frac{d{\mu}}{\mu}{\gamma}_{q}
   \label{sudakov-cq},
   \end{equation}
   \begin{equation}
   S_{V}(t)\ =\
   s(x_{3},p_{3}^{+},1/b_{3})
  +s(\bar{x}_{3},p_{3}^{+},1/b_{3})
  +2{\int}_{1/b_{3}}^{t}\frac{d{\mu}}{\mu}{\gamma}_{q}
   \label{sudakov-ds},
   \end{equation}
  where $I_{0}$, $J_{0}$, $K_{0}$, $Y_{0}$ are Bessel
  functions; the expression of $s(x,Q,1/b)$ can be
  found in Ref.\cite{pqcd1};
  ${\gamma}_{q}$ $=$ $-{\alpha}_{s}/{\pi}$ is the
  quark anomalous dimension;
  ${\alpha}$ and ${\beta}_{i}$ are gluon and quark virtuality,
  respectively, where subscript $i$ on ${\beta}_{i}$ corresponds
  to indices of Fig.\ref{feynman}.
   \begin{eqnarray}
  {\alpha} &=& \bar{x}_{1}^{2}m_{1}^{2}
            +  \bar{x}_{2}^{2}m_{2}^{2}
            -  \bar{x}_{1}\bar{x}_{2}t
   \label{gluon-q2}, \\
  {\beta}_{a} &=& m_{1}^{2} - m_{c}^{2}
               +  \bar{x}_{2}^{2}m_{2}^{2}
               -  \bar{x}_{2}t
   \label{beta-fa}, \\
  {\beta}_{b} &=& m_{2}^{2}
               +  \bar{x}_{1}^{2}m_{1}^{2}
               -  \bar{x}_{1}t
   \label{beta-fb}, \\
  {\beta}_{c} &=& \bar{x}_{1}^{2}m_{1}^{2}
               +  \bar{x}_{2}^{2}m_{2}^{2}
               +  x_{3}^{2}m_{3}^{2}
   \nonumber \\ &-&
                  \bar{x}_{1}\bar{x}_{2}t
               -  \bar{x}_{1}x_{3}u
               +  \bar{x}_{2}x_{3}s
   \label{beta-fc}, \\
  {\beta}_{d} &=& x_{1}^{2}m_{1}^{2}
               +  x_{2}^{2}m_{2}^{2}
               +  x_{3}^{2}m_{3}^{2}
    \nonumber \\ &-&
                  x_{1}x_{2}t
               -  x_{1}x_{3}u
               +  x_{2}x_{3}s
   \label{beta-fd}, \\
   t_{a(b)} &=& {\max}(\sqrt{-{\alpha}},\sqrt{-{\beta}_{a(b)}},1/b_{1},1/b_{2})
   \label{tab}, \\
   t_{c(d)} &=& {\max}(\sqrt{-{\alpha}},\sqrt{{\vert}{\beta}_{c(d)}{\vert}},1/b_{2},1/b_{3})
   \label{tcd}.
   \end{eqnarray}
  \end{appendix}

  
  \end{document}